\documentstyle[prl,eqsecnum,aps,epsf,floats]{revtex}

\begin{document}
\draft
\title{Nature of spin-charge separation}
\author{Z.Y. Weng, D.N. Sheng, and C.S. Ting}
\address{Texas Center for Superconductivity and Department of Physics\\
University of Houston, Houston, TX 77204-5506}
\maketitle

\begin{abstract}
Quasiparticle properties are explored in an effective theory of the $t-J$
model which includes two important components: spin-charge separation and
unrenormalizable phase shift. We show that the phase shift effect indeed
causes the system to be a non-Fermi liquid as conjectured by Anderson on a
general ground. But this phase shift also drastically changes a conventional
perception of quasiparticles in a spin-charge separation state: an injected
hole will remain {\em stable} due to the confinement of spinon and holon by
the phase shift field despite the background is a spinon-holon sea. True 
{\em deconfinement} only happens in the {\em zero-doping} limit where a bare
hole will lose its integrity and decay into holon and spinon elementary
excitations. The Fermi surface structure is completely different in these
two cases, from a large band-structure-like one to four Fermi points in
one-hole case, and we argue that the so-called underdoped regime actually
corresponds to a situation in between, where the ``gap-like'' effect is
amplified further by a microscopic phase separation at low temperature.
Unique properties of the single-electron propagator in both normal and
superconducting states are studied by using the equation of motion method.
We also comment on some of influential ideas proposed in literature related
to the Mott-Hubbard insulator and offer a unified view based on the present
consistent theory.
\end{abstract}


\preprint{} \widetext

\widetext

\section{INTRODUCTION}

High-$T_c$ cuprates are regarded by many as essentially a doped Mott-Hubbard
insulator\cite{anderson}. At half-filling such an insulator is a pure
antiferromagnet with only the spin degrees of freedom not being frozen at
low energy. And a metallic phase with gapless charge degrees of freedom
emerge after holes are added to the filled lower Hubbard band. To
characterize the doped Mott-Hubbard insulator in the metallic regime, two
important ideas were originally introduced by Anderson: Spin-charge
separation\cite{anderson1} and unrenormalizable phase shift effect\cite
{anderson2,anderson3}. The first one is about elementary excitations of such
a system and the second one is responsible for its non-Fermi liquid behavior.

The spin-charge separation idea may be generally stated as the existence of
two independent elementary excitations, charge-neutral spinon and spinless
holon, which carry spin $1/2$ and charge $+e$, respectively. It can be easily
visualized in a short-range resonating-valence-bond (RVB) state\cite{KRS} and
has become a widely used terminology in literature, often with an additional
meaning attached to it. For example, a spin-charge separation may be
mathematically realized in the so-called slave-particle representation\cite
{zou} of the $t-J$ model 
\begin{equation}  \label{slave}
c_{i\sigma}=h^{\dagger}_if_{i\sigma}
\end{equation}
where the no-double-occupancy constraint, reflecting the Hubbard gap in its
extreme limit, is handled by an equality $h^{\dagger}_ih_i+
\sum_{\sigma}f^{\dagger}_{i\sigma}f_{i\sigma}=1$ which commutes with the
Hamiltonian. Here one sees the close relation of the spin-charge separation
and the constraint condition through the counting of the quantum numbers.
But the spin-charge separation also acquires a {\em new} meaning here: If those
holon ($h^{\dagger}_i$) and spinon ($f_{i\sigma}$) fields indeed describe
elementary excitations, the hole (electron) is no longer a stable object and
must decay into a holon-spinon pair once being injected into the system as
shown by Fig. 1(a). This instability of a quasiparticle will be later
referred to as the {\em deconfinement}, in order to distinguish it from the
narrow meaning of the {\em spin-charge separation} about elementary
excitations. We will see later that these two are generally {\em not} the
same thing.

The second idea, the so-called ``unrenormalizable phase shift''\cite
{anderson2,anderson3}, may be described as follows. In the presence of an
upper Hubbard band, adding a hole to the lower Hubbard band could change the
whole Hilbert space due to the on-site Coulomb interaction: The {\em entire}
spectrum of momentum $k$'s may be shifted through the phase shift effect. It
leads to the orthogonality of a bare doped hole state with its true ground
state such that the quasiparticle weight $Z\equiv 0$, the key criterion for
a non-Fermi liquid. In general, it implies 
\begin{equation}  \label{slave2}
c_{i\sigma}=e_{i\sigma} e^{i\Theta_{i}},
\end{equation}
where $e_{i\sigma}$ is related to elementary excitation fields, e.g, $%
h_i^{\dagger}f_{i\sigma}$ in a spin-charge separation framework. Such an
expression means that in order for a bare hole created by $c_{i\sigma}$ to
become low-lying elementary excitations, a {\em many-body} phase shift $%
\Theta_{i}$ must take place in the background. In momentum space, it is easy
to see how such a phase shift changes the Hilbert space by shifting $k$
values. Note that $e_{{\bf k} \sigma} =\sum_{{\bf k^{\prime}}}h^{\dagger}_{%
{\bf k}^{\prime}}f_{{\bf k}+{\bf k}^{\prime}\sigma}$ where ${\bf k}$ and $%
{\bf k}^{\prime}$ belong to the {\em same} set of quantized values (for
example, in a 2D square sample with size $L\times L$, the momentum is
quantize at $k_{\alpha}=\frac{2\pi}{L} n$ under the periodic boundary
condition with $\alpha=x$, $y$ and $n=$ integer). But because of a
nontrivial $\Theta_{i}$, $c_{{\bf k}\sigma}$ and $e_{{\bf k}\sigma}$
generally may no longer be described by the same set of ${\bf k}$'s, or, in
the same Hilbert space, which thus constitutes an essential basis for a
possible non-Fermi liquid.

The one-dimensional (1D) Hubbard model serves a marvelous example in favor
of the decomposition (\ref{slave2}) over (\ref{slave}). The quantitative
value of the phase shift was actually determined by Anderson and Ren\cite
{anderson1,ren} using the Bethe-Ansatz solution\cite{exact} in large $U$
limit, and $Z$ was shown to decay at large sample size with a finite
exponent. An independent path-integral approach\cite{weng1} without using
the Bethe-Ansatz also reaches the same conclusion which supports (\ref
{slave2}) as the {\em correct} decomposition of {\em true} holon and spinon.

Another important property of the Mott-Hubbard insulator, which is
well-known but has not been fully appreciated, is the {\em bosonization} of
the electrons at half-filling. Namely, the {\em fermionic} nature of the
electrons completely disappears and is replaced by a {\em bosonic} one. This
is one of the most peculiar features of the Mott-Hubbard insulator due to
the strong on-site Coulomb interaction. In fact, under the
no-double-occupancy constraint, the $t-J$ model reduces to the Heisenberg
model in this limit. Its ground state for any {\em finite} bipartite lattice
is singlet according to Marshall\cite{marshall} and the wave function is
real and satisfies a {\em trivial} Marshall sign rule as opposed to a much
complicated ``sign problem'' associated to the fermionic statistics in a
conventional fermionic system. This bosonization is the reason behind a very
successful bosonic RVB description of the antiferromagnet: The variational
bosonic RVB wave functions can produce strikingly accurate ground-state
energy\cite{liang,chen} as well as elementary excitation spectrum over the
whole Brillouin zone\cite{chen}. A mean-field bosonic RVB approach\cite{aa},
known as the Schwinger-boson mean-field theory (SBMFT), provides a fairly
accurate and mathematically useful framework for both zero and finite
temperature spin-spin correlations.

Starting from either the slave-boson\cite{weng2} or slave-fermion\cite
{string1} representation, a two-dimensional (2D) version of the
decomposition (\ref{slave2}) has been previously constructed such that the
electron bosonization can be naturally realized at half-filling to restore
the correct antiferromagnetic (AF) correlations. In the 1D case, this
decomposition also recovers the aforementioned spinon-holon decoupling and
reproduces the correct Luttinger liquid behavior\cite{string1}. Even in the
two-leg ladder system where holons and spinons are recombined together to
form quasiparticles in the strong rung case\cite{yu}, a many-body phase
shift field in this decomposition still exists at finite doping, playing a
nontrivial role. Such a decomposition form can be generally written as 
\begin{equation}  \label{mutual}
c_{i\sigma}=h^{\dagger}_ib_{i\sigma}e^{i\hat{\Theta}_{i\sigma}}
\end{equation}
It may be properly called a {\em bosonization} formulation as the holon
operator $h^{\dagger}_{i\sigma}$ and spinon operator $b_{i\sigma}$ are both 
{\em bosonic} fields here. They still satisfy the no-double-occupancy
constraint $h^{\dagger}_ih_i+\sum_{\sigma}b^{\dagger}_{i\sigma}b_{i\sigma}=1$%
. The {\em fermionic} nature of $c_{i\sigma}$ is to be represented through
the phase shift field $\hat{\Theta}_{i\sigma}$ which replaces the
description of Fermi surface patches {\em and} Fermi-surface fluctuations in
the usual bosonization language\cite{haldane,anderson3}. The phase shift
field in 2D is related to a nonlocal vortex-like operator by $e^{i\hat{\Theta%
}_{i\sigma}}\equiv (-\sigma)^ie^{i\Theta_{i\sigma}^{string}}$ (the sign $%
(-\sigma)^i$ keeps track of the Marshall sign just for convenience) with the
vorticity given by\cite{string1} 
\begin{equation}  \label{thetap}
\oint_{\Gamma} d{\bf r}\cdot\nabla \Theta_{\sigma}^{string}= \pm\pi
\sum_{l\in \Gamma} \left[\sum_{\alpha}\alpha n_{l\alpha}^b -1-\sigma
n^h_l\right],
\end{equation}
where $\Gamma$ is an arbitrary closed loop without passing any lattice site
except the sit $i$ and the summation on the r.h.s. of (\ref{thetap}) runs
over lattice sites $l$ within the loop $\Gamma$. Here $n_{l\alpha}^b$ and $%
n_l^h$ are spinon and holon number operators, respectively.

Such a vortex-like phase shift originates from the fact that a doped hole
moving on an AF spin background will always pick up sequential $+$ and $-$
signs, $(+1)\times (-1)\times (-1)\times ...$, first identified in Ref.%
\onlinecite{string}. These signs come from the Marshall signs hidden in the
AF background which are scrambled by the hopping of the doped hole on its
path, determined by simply counting the index $\sigma $ of each spin
exchanged with the hole during its hopping\cite{string}. The significance of
such a phase string is that it represents the {\it sole} source to generate
phase frustrations in the $t-J$ model (at finite doping, the only additional
signs coming from the fermionic statistics of doped holes in the original
slave-fermion representation are also counted in). Namely, the ground-state
wave function would become real and there should be no ``sign problem'' only
if such a phase string effect is absent (like in the zero-doping case). The 
phase
shift field in (\ref{mutual}) precisely keeps track of such a phase string
effect\cite{string1} and therefore can be considered to be a general
consequence of the $t-J$ model.

The decomposition (\ref{mutual}) {\em defines} a unique spin-charge
separation theory where the relation between the physical electron operator
and the internal elementary excitations, holon and spinon, is explicitly
given. The thermodynamic properties will be obtained in terms of the energy
spectra of holon and spinon fields, while the physically observable
quantities will be determined based on (\ref{mutual}) where the singular
phase shift field with vorticities is to play a very essential role in
contrast to the conventional spin-charge separation theories in the
slave-particle decompositions of (\ref{slave}).

Note that the total vorticity of (\ref{thetap}) is always equal to $2\pi
\times integer $ due to the no-double-occupancy constraint such that the
phase shift factor $e^{i\hat{\Theta}_{i\sigma}}$ is single-valued. Such a
phase shift field will play different roles in different channels. For
example, in the spin channel one has $S^{+}_i=
b^{\dagger}_{i\uparrow}b_{i\downarrow} (-1)^i
e^{-i\left[\Theta_{i\uparrow}^{string}-\Theta_{i\downarrow}^{string}\right]}$
where the total vorticity 
\begin{equation}  \label{spinp}
-\oint_{\Gamma}d{\bf r}\cdot \nabla \left(
\Theta_{\uparrow}^{string}-\Theta_{\downarrow}^{string}\right)= \pm 2\pi
\sum_{l\in \Gamma} n^h_l.
\end{equation}
It obviously vanishes at $\delta\rightarrow 0$ ($\delta$ is the doping
concentration) so that the aforementioned bosonization is naturally
realized. And at finite-doping, the vorticity shown in (\ref{spinp})
reflects the recovered fermionic effect and is responsible for a
doping-dependent incommensurate momentum structure\cite{string2} in the
dynamic spin susceptibility function which provides a unique reconciliation
of neutron scattering and NMR measurements in the cuprates. On the other
hand, in the {\em singlet} pairing channel, the phase shift field appearing
in the local pairing operator will contribute to a vorticity 
\begin{equation}  \label{pairp}
\oint_{\Gamma}d{\bf r}\cdot \nabla \left(
\Theta_{\uparrow}^{string}+\Theta_{\downarrow}^{string}\right)= \pm
2\pi\sum_{l\in \Gamma} \left[\sum_{\alpha}\alpha n_{l\alpha}^b -1 \right],
\end{equation}
which decides the {\em phase coherence} of Cooper pairs. According to (\ref
{pairp}), besides a trivial $2\pi$ flux quantum per site, each spinon also
carries a fictitious $(\pm)2\pi$ flux tube. To achieve the phase coherence
or superconducting condensation, $\uparrow$ and $\downarrow$ spinons have to
be {\em paired} off to remove the vorticities associated with individual
spinons in (\ref{pairp}), which then connects\cite{string3,string4} $T_c$ to
a characteristic spinon energy scale, in consistency with the experimental
result of cuprate superconductors and resolving the issue why $T_c$ is too
high in usual RVB theories.

The purpose of the present work is to explore the consequences of the
bosonization decomposition (\ref{mutual}) in 2D {\em quasiparticle} channel.
First of all, we show that the phase shift field indeed causes the
quasiparticle weight $Z$ to vanish. Namely, this spin-charge separation
state {\em is} a 2D non-Fermi liquid, a fact almost trivial in such a
particular formulation. A surprising ``by-product'' of this phase shift
field is that it also plays a role of {\em confinement force} to ``glue''
spinon and holon constituents together inside a quasiparticle, as
illustrated in Fig. 1(b). In other words, a hole injected into this system
generally does {\em not} break up into spinon-holon elementary particles,
even though the background is a spinon-holon sea. Such a quasiparticle may
be regarded as a spinon-holon bound state or more properly a {\em collective}
mode but will generally remain {\em incoherent} due to the same phase shift
field.
\begin{figure}[b!]
\epsfxsize=5.0 cm
\centerline{\epsffile{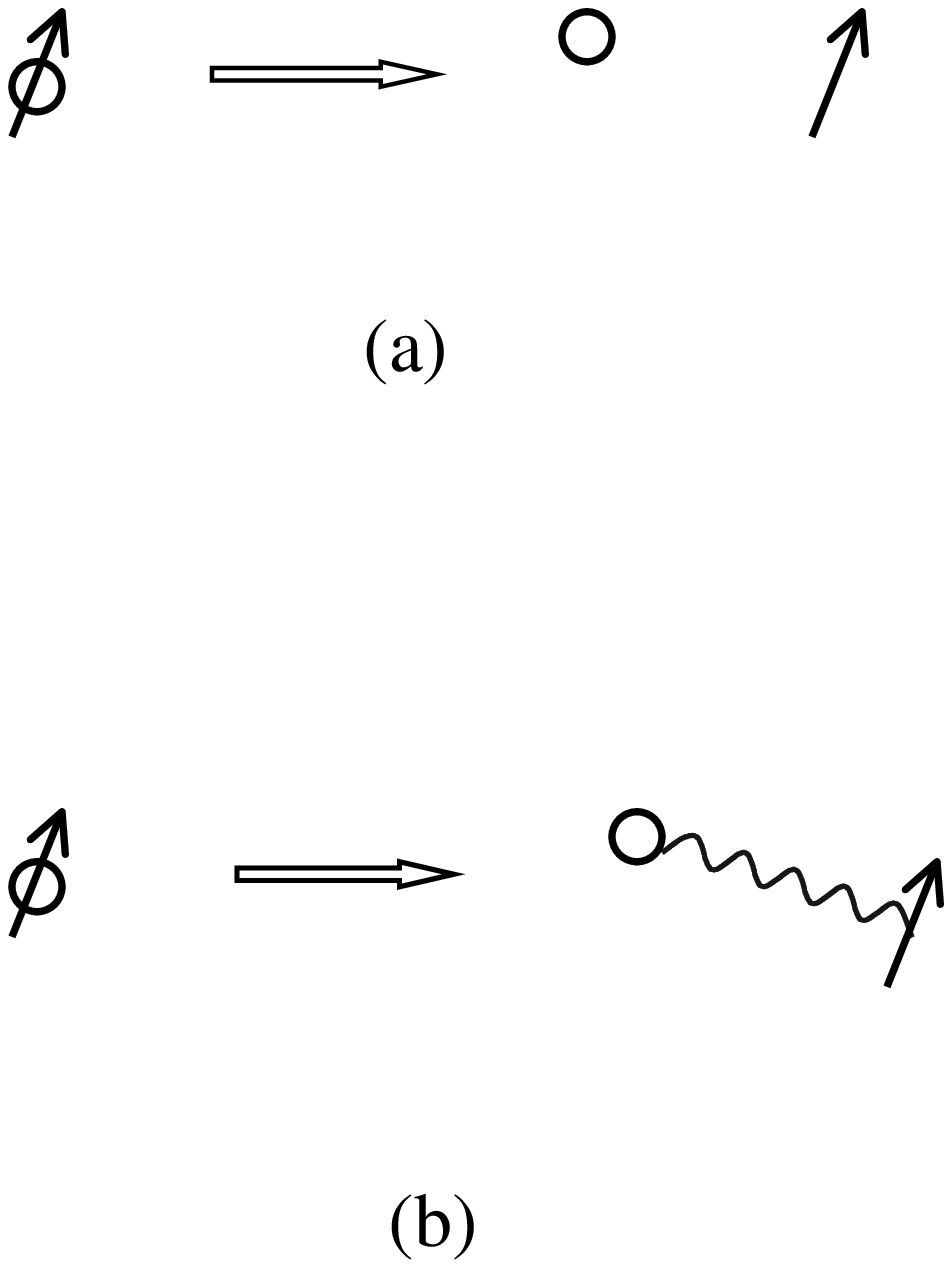}}
\vspace{2mm}
\caption{Schematical illustration of the quasiparticle deconfinement (a) and 
confinement (b) due to the phase shift field} 
\label{fig:1}
\end{figure}

Due to the confinement, an equation-of-motion description of the
quasiparticle excitation is constructed, in which the dominant
``scattering'' process is described as the ``virtual'' decay of the
quasiparticle into holon-spinon composite. In the superconducting phase, the
composite nature of the quasiparticle predicts a unique non-BCS structure
for the single-electron Green's function which is consistent with the
experimental measurements. In particular, we find the restoration of the
quasiparticle coherence with regard to the incoherence in the normal state.

A true deconfinement or instability of the quasiparticle only happens in the
zero-doping limit where an injected hole indeed can decay into a holon and
spinon pair, which provides\cite{1hole} a consistent explanation of
angle-resolved photoemission spectroscopy (ARPES) measurements\cite{arpes1}.
The contrast of a large band-structure-like Fermi surface in the confinement
phase to the four Fermi-points in the deconfinement phase at zero-doping
limit may provide a unique explanation for the ARPES experimental
measurements in cuprate superconductors. In a weakly-doped regime, a
``partial'' deconfinement of the quasiparticle between full-blown
deconfinement and confinement will be reflected in the single-electron
Green's function which may explain the ``spin gap'' phenomenon\cite
{arpes2}.

The remainder of the paper is organized as follows. In Sec. II, we first
briefly review the effective spin-charge separation theory based on the
decomposition (\ref{mutual}). Then in Sec. II B, we show that the phase
shift field leads to $Z=0$, i.e., a non-Fermi liquid state. In Sec. II C, we
demonstrate how the phase shift field causes the confinement of the holon
and spinon within a quasiparticle except in the zero-doping limit. In Sec,
II D, we study the single-electron propagator in both normal and
superconducting states based on an equation-of-motion approach. We then
discuss an underdoped case as a crossover regime from a Fermi-point
structure in one-hole case with the holon-spinon deconfinement to a large
Fermi-surface in the confinement case. Finally, Sec. III is devoted to
discuss some of the most influential ideas proposed in literature related to
the Mott-Hubbard insulator and high-$T_c$ cuprates and offer a unified view
based on the present consistent theory.

\section{Properties of a quasiparticle in spin-charge separation state}

\subsection{Effective spin-charge separation theory}

The decomposition (\ref{mutual}) determines an effective spin-charge
separation theory of the $t-J$ model in which spinon and holon fields
constitute the elementary particles. Before proceeding to the discussion of
the quasiparticle properties in next subsections, we first briefly review
some basic features of this theory based on Refs.\onlinecite{string1,string3}.

In the operator formalism, the phase shift field $\Theta_{i\sigma}^{string}$
satisfying (\ref{thetap}) can be explicitly written down in a specific gauge
as follows\cite{string1} 
\begin{equation}  \label{theta}
\Theta_{i\sigma}^{string}\equiv \frac i 2 \left[\Phi^b_i-\sigma
\Phi_i^h\right],
\end{equation}
where 
\begin{equation}  \label{phib}
\Phi^{b}_i= \sum_{l\neq i} \theta_i(l)\left(\sum_{\alpha}\alpha
n_{l\alpha}^b -1\right),
\end{equation}
and 
\begin{equation}  \label{phih}
\Phi_i^h= \sum_{l\neq i}\theta_i(l) n_l^h.
\end{equation}
Here $\theta_i(l)$ is defined as an angle 
\begin{equation}  \label{angle}
\theta_i(l)=\mbox{Im ln $(z_i-z_l)$}
\end{equation}
with $z_i=x_i+iy_i$ representing the complex coordinate of a lattice site $i$.

In 2D, an effective Hamiltonian based on the decomposition (\ref{mutual})
after a generalized mean-field decoupling\cite{string3} in the $t-J$ model
can be written down: 
\begin{equation}
H_{eff}=H_h+H_s
\end{equation}
where the holon Hamiltonian 
\begin{equation}  \label{hh}
H_h= -t_h \sum_{\langle ij\rangle}\left(e^{iA_{ij}^f}\right)h^{\dagger}_ih_j
+ H.c.
\end{equation}
and the spinon Hamiltonian 
\begin{eqnarray}  \label{hs}
H_s =& -& J_s \sum_{\langle ij\rangle \sigma }\left[ \left(e^{i\sigma
A_{ij}^h}\right)b_{i\sigma}^{\dagger}b^{\dagger}_{j-\sigma}+ H.c.\right] 
\nonumber \\
& - &\sum_{ij \sigma } J_{ij}^s \left(e^{i\sigma
A_{ij}^h}\right)b_{i\sigma}^{\dagger}b_{j\sigma},
\end{eqnarray}
with $t_h\sim t$, $J_s\sim J$. In the second term of $H_s$, $J^s_{ij}\sim \delta 
t \neq 0$ only for $i $ and $j$ on the same sublattice sites, which originates 
from $H_t$ where a phase shift occurs\cite{string3} in the spinon mean-field 
wavefunction and results in the same-sublattice hopping of spinons. The lattice 
gauge fields $A_{ij}^f$ and $A_{ij}^h$ in the specific gauge choice of 
(\ref{phib}) and (\ref{phih}) are given by 
\begin{equation}  \label{eaf}
A_{ij}^f =\frac 1 2 \sum_{l\neq i, j}\left[\theta_i(l)-\theta_j(l)
\right]\left(\sum_{\sigma}\sigma n_{l\sigma}^b-1\right)\equiv
A_{ij}^s-\phi_{ij}^0,
\end{equation}
and 
\begin{equation}  \label{eah}
A_{ij}^h=\frac 1 2 \sum_{l\neq i,
j}\left[\theta_i(l)-\theta_j(l)\right]n_l^h.
\end{equation}
In general, $A_{ij}^s$ and $A_{ij}^h$ can be regarded as ``mutual''
Chern-Simons lattice gauge fields as, for example, $A_{ij}^h$ is determined
by the density distribution of holons but only seen by spinons.

The above effective theory is based on a RVB pairing order parameter\cite
{string3} 
\begin{equation}  \label{deltas}
\Delta^s=\sum_{\sigma}\left\langle e^{-i\sigma
A^h_{ij}}b_{i\sigma}b_{j-\sigma}\right\rangle,
\end{equation}
which in zero-doping limit $\delta\rightarrow 0$ reduces to the well-known
bosonic RVB order parameter\cite{aa} as $A^h_{ij}=0$. And $H_{eff}$ recovers
the Schwinger-boson mean-field Hamiltonian\cite{aa} of the Heisenberg model.
So this theory can well describe AF correlations at half-filling. At finite
doping, the ``mutual'' Chern-Simons gauge fields, $A_{ij}^f$ and $A^h_{ij}$,
will play important roles in shaping superconductivity, magnetic and
transport properties, and some very interesting similarities with cuprate
superconductors have been discussed based on this model\cite{string3}. In 
contrast to the slave-fermion approach\cite{slave-fermion}, $\Delta^s$ remains
the only order parameter at finite doping, controlling the short-range spin-spin
correlations as $\langle {\bf S}_i\cdot{\bf S}_j\rangle=-1/2 |\Delta^s|^2$ for
nearest-neighboring $i$ and $j$.
It is noted that due to the presence of the RVB pairing (\ref{deltas}), the
conventional gauge fluctuations\cite{gauge1,gauge2} are suppressed as the
gauge invariance, $h^{\dagger}_ib_{i\sigma}=[h^{\dagger}_ie^{i
\theta_i}][b_{i\sigma}e^{- i\theta_i}]$, is apparently broken by $\Delta^s$.
Here spinons no longer contribute to transport and are really charge-neutral
particles.

In the ground state of the uniform-phase solution (Ref.\onlinecite{string3}%
), $A_{ij}^f$ and $A_{ij}^h$ both become simplified: $A_{ij}^f$ simply
describes a $\pi$-flux per plaquette: $\sum_{\Box} A_{ij}^f\approx
-\sum_{\Box} \phi^0_{ij}=-\pi$ since $A^s_{ij}$ is suppressed due to the
spinon pairing in the ground state; $A_{ij}^h$ describes a uniform flux $%
\sum_{\Box} \bar{A}_{ij}^f=\pi\delta$ due to the Bose condensation of
holons. Self-consistently, $H_h$ [(\ref{hh})] determines a Bose-condensed
ground state of holons where the $\pi$ flux produced by $A_{ij}^f$ merely
enlarges the effective mass near the band edge by $\sqrt{2}$. On the other
hand, $\bar{A}^h_{ij}$ in $H_s$ [(\ref{hs})] leads to a ``resonance-like'' energy structure in the spinon spectrum and the 
corresponding dynamic spin susceptibility function at the AF vector 
($\pi$, $\pi$)  is illustrated in Fig. 2. Note that $E_g$ in Fig. 2 is twice bigger
than the corresponding spinon energy $E_s$. The doping dependence of $E_g$
is shown in the inset. Finally, the
superconducting order parameter $\Delta^{SC}_{ij}=\langle
\sum_{\sigma}\sigma c_{i\sigma}c_{j-\sigma}\rangle$ has a finite value (for
nearest-neighboring $i$ and $j$) in the ground state\cite{string3} 
\begin{equation}  \label{sc}
\Delta^{SC}_{ij}= \Delta^s (-1)^i \langle h^{\dagger}_i e^{\frac i 2
(\Phi^b_i+\Phi^b_j)}h^{\dagger}_j\rangle\neq 0,
\end{equation}
due to the Bose condensation of holons as well as the pairing of spinons
which leads to $\Delta^s\neq 0$ and the vortex-antivortex confinement in $e^{%
\frac i 2 (\Phi^b_i+\Phi^b_j)}$. Thus the ground state is always
superconducting-condensed with a pairing symmetry of d-wave-like\cite
{string3}.

Besides the above uniform ground state, possible non-uniform solutions
characterized by the coexistence of the Bose condensations of holons and
spinons have been also discussed in Ref.\onlinecite{string3} where the
spinon spectrum has only a pseudo-gap. In any case, the ground state can be
regarded as a spinon-holon sea, and low-lying elementary excitations are
described in terms of spinons and holons. What we are mainly interested in
this paper is to answer the question how a hole (electron) as a composite of
spinon and holon behaves in this spin-charge separation state. This is one
of the most fundamental questions not only because it can be directly tested
in an ARPES measurement, but also because it will make crucial distinction
between a conventional Fermi liquid and a non-Fermi liquid. Let us begin
with the question: If this spin-charge separation state is a non-Fermi
liquid. 
\begin{figure}[ht!]
\epsfxsize=8.0 cm
\centerline{\epsffile{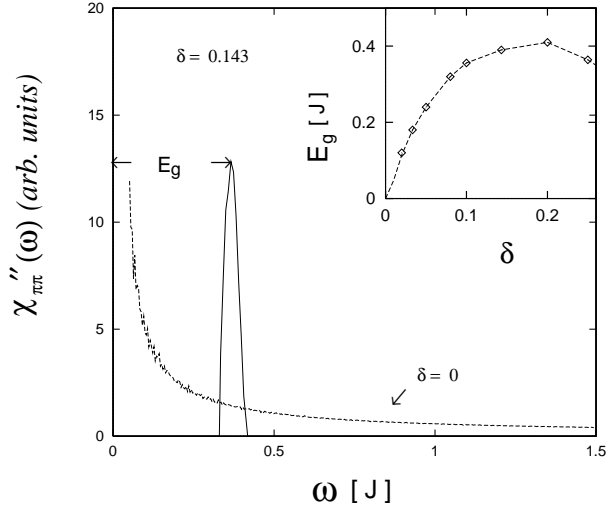}}
\vspace{2mm}
\caption{Dynamical spin susceptibility at the AF vector ($\pi$, $\pi$) in the
uniform phase for $\delta=0$ and $0.143$, respectively. The inset: the 
doping dependence of the characteristic spin ``resonance-like'' energy 
$E_g=2E_s$ ($E_s$ denotes the corresponding spinon energy).  } 

\label{fig:2} 
\end{figure}

\subsection{Non-Fermi liquid with $Z=0$}

The definition of the quasiparticle weight $Z_{{\bf k}}$ at momentum ${\bf k}
$ is given by $Z_{{\bf k}}=|\langle \Psi_G(N_e-1)|c_{{\bf k}%
\sigma}|\Psi_G(N_e)\rangle|^2$, and it measures the overlap of a bare hole
state at momentum ${\bf k}$, created by $c_{{\bf k}\sigma}$ in the ground
state of $N_e$ electrons, with the ground state of $N_e-1$ electrons. For a
Fermi liquid state, one always has $Z_{k_f}\neq 0$ at the Fermi momentum $k_f
$. If $Z_{{\bf k}}=0$ for any ${\bf k}$, then the system is a non-Fermi
liquid by definition. In the following we will show that the bare hole state 
$c_{{\bf k}\sigma}|\Psi_G(N_e)\rangle$ acquires an ``angular'' momentum due
to the vorticities in $e^{\hat{\Theta}_{i\sigma}}$ in the decomposition (\ref
{mutual}). Due to such a distinct symmetry, it is always orthogonal to the
ground state $|\Psi_G(N_e-1)\rangle$, leading to $Z\equiv 0$.

One can construct a ``rotational'' operation by making a transformation 
\begin{equation}  \label{transf}
\theta_i(l)\rightarrow \theta_i(l)+\phi.
\end{equation}
It corresponds to a simple change of reference axis for the angle function $%
\theta_i(l)$ defined in (\ref{angle}). It is easy to see that the
Hamiltonians (\ref{hh}) and (\ref{hs}) are invariant since the gauge fields,
in which $\theta_i(l)$ appears, are obviously not changed: $A_{ij}^{h,
f}\rightarrow A_{ij}^{h, f}$. Both the ground state $|\Psi_G\rangle$ as well
as single-valued $h_i^{\dagger}$ and $b_{i\sigma}$ fields are apparently
independent of $\phi$. But a bare hole state will change under the
transformation (\ref{transf}) as follows: 
\begin{equation}  \label{ctransf}
c_{i\sigma}|\Psi_G\rangle \rightarrow e^{i\phi P_i^{\sigma}}\times
c_{i\sigma}|\Psi_G\rangle
\end{equation}
due to the phase shift factor $e^{i\hat{\Theta}_{i\sigma}}$ with $%
P_i^{\sigma}=S^z -{\sigma}{N^h}/2-(N-1-\sigma)/2$ which remains an integer
for a bipartite lattice ($S^z$ and $N^h$ denote total spin and hole numbers,
respectively, and N is the lattice size). This implies that $%
c_{i\sigma}|\Psi_G\rangle$ indeed has a nontrivial ``angular'' momentum in
contrast to $|\Psi_G\rangle$ which carries none.

It is probably more transparent to see the origin of the angular momentum if
one rewrites, for example, 
\begin{equation}  \label{z}
e^{i\hat{\Theta}_{i\downarrow}}=\prod_{l\neq
i}\left(z_i-z_{l\uparrow}\right)^{1/2}\prod_{l\neq
i}\left(z_i^*-z_{l\downarrow}^*\right)^{1/2}\prod_{l\neq
i}\left(z_i-z_{lh}\right)^{1/2}\prod_{l\neq
i}\left(z_i^*-z_{l}^*\right)^{1/2}\times F_i
\end{equation}
where $z_{l\uparrow}$, $z_{l\downarrow}$, and $z_{lh}$ denote the complex
coordinates of $\uparrow$, $\downarrow$ spinons, and holons, respectively.
And $F_i=\prod_{l\neq i}\left|z_i-z_{l}\right|$ is a constant (which is
obtained by using the no-double-occupancy constraint). It is important to
note that despite the fractional (``semion'') exponents of $1/2$ in (\ref{z}%
), it can be directly verified that the phase shift field $e^{i\hat{\Theta}%
_{i\downarrow}}$ remains {\em single-valued} under the no-double-occupancy
constraint. Generally the vortex field (\ref{z}) introduces an extra angular
momentum which can be easily identified as 
\begin{equation}
l=S^z+\frac {N^h}{2}.
\end{equation}
Here $l$ is always an integer. Then one has 
\begin{equation}  \label{z1}
\langle \Psi_G(N_e-1)|c_{i\sigma}|\Psi_G(N_e)\rangle=0
\end{equation}
due to the {\em orthogonal} condition\cite{remark0} as $l\neq 0$ [$S^z=O(1)$%
, $N^h=O(N)$ at finite doping] for $c_{i\sigma}|\Psi_G\rangle$. By extending
the same argument, one can quickly see that the bare hole state $%
c_{i\sigma}|\Psi_G(N_e)\rangle$ has no overlap not only with $%
|\Psi_G(N_e-1)\rangle$ but also with all the elementary excitations composed
of simple holons and spinons with $l=0$. So $c_{i\sigma}$ is more like a
creation operator of a ``collective'' mode whose quantum number $l$ is
different from a simple spinon-holon pair.

\subsection{Quasiparticle: Spinon-holon confinement}

The difference in symmetry between a quasiparticle and a holon-spinon pair
implies that the former cannot simply decay into the latter even though they
share the same quantum numbers of charge and spin. In this section, we
demonstrate that generally the holon and spinon constituents will be {\em %
confined} by the phase shift field within a quasiparticle although the
background is a spinon-holon sea.

Intuitively such a confinement is easy to understand: If the holon and
spinon constituents inside a quasiparticle could move away independently by
themselves, as schematically shown in Fig. 3, the vortex-phase-shift field $%
e^{i\hat{\Theta}_{i\sigma}}$ left behind would cost a logarithmically {\em %
divergent} energy as to be shown below. But a quasiparticle state $%
c_{i\sigma}|\Psi_G\rangle$ as a local excitation should only cost a finite
energy relative to the ground state energy. Such a discrepancy can be
reconciled only if the holon and spinon constituents no longer behave as
free elementary excitations: They have to absorb the effect of the
vortex-like phase shift and by doing so make themselves bound together.

Let us consider $|\Psi^{\prime}\rangle\equiv e^{i\hat{\Theta}%
_{i\sigma}}|\Psi_G\rangle$ and compute the energy cost for the vortex-like
phase shift: 
\begin{equation}  \label{E'}
\langle \Psi^{\prime}|H_{eff}|\Psi^{\prime}\rangle - \langle
\Psi_G|H_{eff}|\Psi_G\rangle.
\end{equation}
We first focus on the contribution from the holon part $H_h$ [(\ref{hh})].
Define $E_G^h=\langle \Psi_G|H_h|\Psi_G\rangle$. One has $-t_h\langle
\Psi_G|h_l^{\dagger}h_me^{iA_{lm}^f}|\Psi_G\rangle= E^h/4N$ for any nearest
neighbor link (lm) due to the translational symmetry. Then a straightforward
manipulation leads to 
\begin{equation}  \label{Eh'}
\langle \Psi^{\prime}|H_{h}|\Psi^{\prime}\rangle - E_G^h=-\frac{E^h_G}{2N}%
\sum_{\langle lm\rangle}\left\{1-\cos
\left[\theta_i(l)-\theta_i(m)\right]/2\right\}.
\end{equation}
Notice that if the (lm) link (say, along $\hat{x}$-direction) is far away
from the site $i$, then one has $|\theta_i(l)- \theta_i(m)|\rightarrow a
|\sin \theta|/r$ where $r$ denotes the distance between the center of the
link and the site $i$ and $\theta$ is the azimuth angle. Then it is easy to
see that the summation over those links on the r.h.s. of (\ref{Eh'}) will
contribute as $\int r dr d\theta \sin ^2\theta /r^2 \propto \ln R $ ($R$
denotes the sample size). Namely, the vortex-like phase shift will cost a
logarithmically diverged energy if it is left alone. Similar logarithmically
divergent energy contributed by $H_s$ [(\ref{hs})] can be also obtained. It
should be noted that the same conclusion still holds if one replace $H_{eff}$
by the exact $t-J$ model in the representation of (\ref{mutual}) (Ref.%
\onlinecite{string1}).
\begin{figure}[ht!]
\epsfxsize=8.0 cm
\centerline{\epsffile{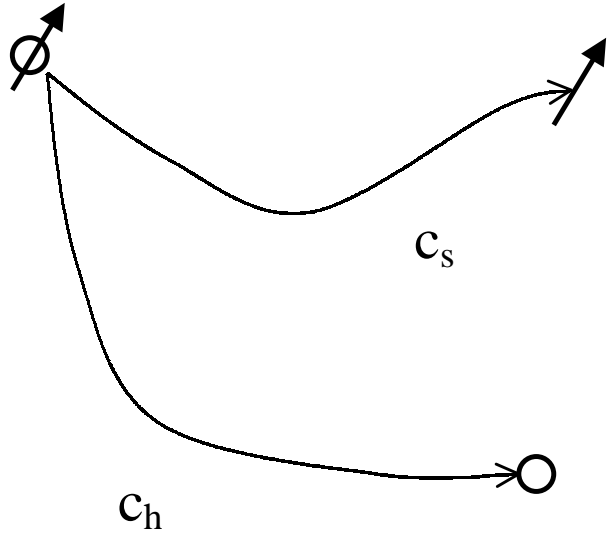}}
\vspace{2mm}
\caption{Schematical illustration of the case when a quasiparticle decays 
into spinon and holon constituents.} 
\label{fig:3}
\end{figure}

Hence the vortex-like phase shift field has to be absorbed by the holon and
spinon fields in order to keep the quasiparticle energy finite. In the
following, let us illustrate how this will happen. We first use the vortex
phase $\frac{1}{2}\Phi^b_i$ in (\ref{theta}) as an example. Let us write
down the following identity 
\begin{equation}  \label{lineh}
e^{i\frac{1}{2}\Phi^b_{i}}= \left(e^{i \sum_{c_h}A^f} e^{iK^b(c_h)}\right)
e^{i\frac{1}{2}\Phi^b_{j}},
\end{equation}
in which 
\begin{equation}
\sum_{c_h}A^f\equiv \sum_{s}A_{m_sm_{s+1}}^f
\end{equation}
where $m_0=i$, $m_1$, ..., $m_{k_{c_h}}\equiv j$ are sequential lattice
sites on an arbitrary path $c_h$ connecting $i$ and $j$. And $K^b({c_h}%
)\equiv \frac{1}{2}\sum_{s}\left[\theta_{m_{s-1}}(m_s)-\theta_{m_{s+1}}(m_s)%
\right]\left(\sum_{\alpha}\alpha n_{m_s\alpha}^b-1\right)$ which is a
string-like field only involving spinons on the path $c_h$. By contrast, the
line summation $\sum_{c_h} A^f$ is contributed by spinons from the whole
system nonlocally. Note that, according to $H_h$ [(\ref{hh})], holons see
the gauge field $A^f$ in the Hamiltonian and if a holon moves from site $i$
to $j$ via the same path $c_h$ it should acquire a phase factor $%
e^{-i\sum_{c_h} A^f}$ which can exactly compensate the similar phase in (\ref
{lineh}). In other words, if the vortex phase factor $e^{i\frac{1}{2}%
\Phi^b_{i}}$ is bound to a holon to form a new composite object $\tilde{h}%
^{\dagger}_i=h^{\dagger}_ie^{i\frac 1 2 \Phi^b_i}$, then there will be no
more vortex effect as it moves on the path $c_h$ shown in Fig. 3, except for
a phase string field $K^h(c_h)$ left on its path, and the new object should
cost only a finite energy.

Similarly, for the vortex field $-\frac{\sigma}{2}\Phi_i^h$ in (\ref{theta})
one can rewrite 
\begin{equation}  \label{lines}
e^{-i\frac{\sigma}{2}\Phi^h_{i}}= \left(e^{-i\sigma \sum_{c_s}A^h}
e^{-i\sigma K^h(c_s)}\right) e^{-i\frac{\sigma}{2}\Phi^h_{j^{\prime}}},
\end{equation}
in which 
\begin{equation}
\sum_{c_s}A^h\equiv \sum_{s}A_{m_sm_{s+1}}^h
\end{equation}
where the line summation runs over a sequential lattice sites on an
arbitrary path $c_s$ connecting $i$ and $j^{\prime}$ shown in Fig. 3. And
then we can similarly see that the spinon constituent of the quasiparticle
also has to be bound to $e^{-i\frac{\sigma}{2}\Phi^h_{}}$ in order to
compensate the logarithmically divergent energy cost by the vortex
structure, and the new composite will only leave a phase string behind given
by $K^h({c_s})\equiv \frac{1}{2}\sum_{s}\left[\theta_{m_{s-1}}(m_s)-%
\theta_{m_{s+1}}(m_s)\right]n_{m_s}^h$.

However, there is one problem in the above argument about the absorption of
the vortex phase factors $e^{i\frac{1}{2}\Phi^b_i}$ and $e^{i\frac{\sigma}{2}%
\Phi^h_i}$ by the holon and spinon constituents. Namely, these two phase
factors are not single-valued except in the zero doping limit. In fact, only
the total phase factor $e^{i\hat\Theta_{i\sigma}}$ is always well-defined
and single-valued as mentioned before. It thus means that both holon and
spinon constituents have to be bound to the total phase shift fields
together to eliminate the divergent energy while maintain single-valued.
There is another way to see this. Note that $\theta_{m_{s-1}}(m_s)-%
\theta_{m_{s+1}}(m_s)$ describes the angle between the nearest-neighboring
links ($m_{s-1}$, $m_s$) and ($m_{s+1}$, $m_s$), it can have an uncertainty
by $\pm 2\pi\times $integer, and it is easy to see that the phase string
factors $e^{iK^b(c_h)}$ and $e^{-i\sigma K^h(c_s)}$ in (\ref{lineh}) and (%
\ref{lines}) are not well-defined by themselves as they are multi-valued
except at $\delta=0$. On the other hand, if one chooses $c_h=c_s=c$ in Fig.
3, the mathematical ambiguity is eliminated in the total phase string field 
\begin{eqnarray}  \label{kc}
K_{\sigma}(c)&\equiv & K^b(c)-\sigma K^h(c)  \nonumber \\
&= &\sum_{s=1}^{k_{c}}\left[\theta_{m_s-1}(m_s)-\theta_{m_s+1}(m_s)\right] 
\frac 1 2 \left(\sum_{\alpha}\alpha n_{m_s\alpha}^b-1-\sigma
n^h_{m_s}\right),
\end{eqnarray}
since by using the no-double-occupancy constraint, one can show $\frac 1 2
\left(\sum_{\alpha}\alpha n_{m_s\alpha}^b-1-\sigma n^h_{m_s}\right)= -\frac{%
1+\sigma}{2} +\sigma n^b_{m_s\sigma}$ which is an integer such that $%
e^{iK_{\sigma}(c)}$ remains single-valued.

Physically, it is because a {\em fermionic} quasiparticle may not decay into
two {\em bosonic} holon and spinon elementary excitations in 2D. The only
exception is in the {\em zero-doping} limit. We have pointed out in
Introduction that at half-filling the ``fermionic'' nature of the electrons
essentially disappears and is replaced by a ``bosonic'' one due to the
no-double-occupancy constraint. Then it is not surprising that in the
one-hole doped case which is adjacent to the half-filling, the deconfinement
of holon-spinon can happen as a result of the electron ``bosonization''.
Indeed, in the zero-doping limit $\Phi^h_{i}$ defined in the gauge (\ref
{phih}) {\em vanishes}. Without $\Phi^h_i$, the original reason for
inseparable $e^{-i\frac {\sigma} 2 \Phi^h_i}$ and $e^{i\frac 1 2
\Phi^b_{i\sigma}}$ in $e^{i\hat{\Theta}_{i\sigma}}$ is no longer present: In
this case, the phase shift field $e^{i\hat{\Theta}_{i\sigma}}$ reduces to $%
e^{ i\frac 1 2 \Phi^b_i}$ which itself becomes well-defined, and can solely
accompany the holon during the propagation. As for the spinon part, $H_s$ in
(\ref{hs}) reduces to the well-known SBMFT Hamiltonian with $A^h=0$ and the
corresponding line summation $\sum_c A^h$ is also absent in the propagator.
Without leading to the multi-value ambiguity, the quasiparticle will break
into a spinon and a composite of holon-vortex phase which can propagate
independently. More discussions of the one-hole problem can be found in the
Sec. II D3.

How a quasiparticle behaves in a spinon-holon sea as a single entity at
finite doping will be the subject of discussion in the next subsection. In
the following we will make several remarks on some implications of the
confinement before concluding the present subsection. First of all, we note
that a quasiparticle generally remains an incoherent excitation in contrast
to the coherent spinons and holons and we assume that it will not contribute
significantly to either thermodynamic and dynamic properties. In the
equal-time limit $t=0^-$, the single-electron propagator can be expressed as 
\begin{equation}  \label{prop1}
G_e(i,j; 0^-)=i(-\sigma)^{i-j}\left\langle \left(b^{\dagger}_{j\sigma}e^{i%
\frac {\sigma}{2}\sum_c A^h}b_{i\sigma}\right)\left(h_j e^{-i \sum_c A^f }
h_i^{\dagger}\right)e^{iK_{\sigma}(c)}\right\rangle.
\end{equation}
At finite $t$, temporal components have to be added to the line summations, $%
\sum_c A^h$ and $\sum_c A^f$, as well as in the phase string field $%
K_{\sigma}(c)$ above. Even though mathematically the path $c$ can be chosen
arbitrarily in (\ref{prop1}), a natural choice is for $c$ to coincide with
the real path of the quasiparticle such that the line summations can be
precisely compensated by the phases picked up by the holon and spinon
constituents as mentioned above. In this case, all the singular phase effect
will be tracked by $e^{iK_{\sigma}(c)}$ which is nothing but the
previously-identified phase string effect\cite{string}, where it has been
shown that the phase string effect is nonrepairable and represents the
dominant phase interference at low energy. Physically, it reflects the {\em %
fermionic} exchange relation between the quasiparticle under consideration
and those electrons in the background. Such a phase string field
accompanying the propagation of the quasiparticle is a many-body operator in
terms of elementary holon and spinon fields. Even in the one hole case, such
a phase string effect results in the incoherency of the quasiparticle as has
been discussed in Ref. \onlinecite{1hole}.

One may also see how a Fermi-surface structure is generated from the phase
shift $\hat{\Theta}_{i\sigma}$ in some limits. For example, in the 1D case
(where $A_{ij}^{h, f}=0$\cite{string1}), since one may always define $%
\theta_{m_s-1}(m_s)-\theta_{m_s+1}(m_s)= \pm \pi$, the phase string factor $%
e^{iK_{\sigma}(c)}$ in (\ref{prop1}) can be written as $(-\sigma)^{i-j}e^{i%
\sigma k_f(x_i-x_j)}e^{i\delta k_f^{\sigma}(x_i-x_j) }$ which produces the
1D Fermi surface at $k_f=\pm\pi (1-\delta)/2$ (here $\delta k_f^{\sigma}$
denotes Fermi-surface fluctuations with $\langle \delta k_f^{\sigma}\rangle=0
$ which is crucial to the Luttinger liquid behavior\cite{string1}). In the
2D one-hole case, $\hat{\Theta}_{i\sigma}$ also leads to a ``remnant'' Fermi
surface structure in the equal time limit while gives rise to four Fermi
points ${\bf k}_0$ {\em at low energy} as discussed in Ref.\onlinecite{1hole}%
. At finite doping, the doping-dependent incommensurate peaks in the dynamic
spin susceptibility function has been also related to such a phase shift
field\cite{string2}. In general, the Luttinger volume theorem may even be
understood based on $e^{iK_{\sigma}(c)}$ as it involves the counting of the
background electron numbers. Nevertheless, the {\em precise} Fermi-surface
topology will not be solely determined by the phase shift field in 2D and
one must take into account of the dynamic effect.

Finally, a stable but incoherent quasiparticle excitation in which a pair of
holon and spinon are confined means that a photoemission experiment, in
which such a quasiparticle excitation can be created through ``knocking
out'' an electron by a photon, does not directly probe the {\em intrinsic}
information of coherent elementary excitations anymore, and the
energy-momentum structure of the single-electron Green's function is no
longer a basis as fundamental and useful as in the case of conventional
Fermi liquid metals to understand superconductivity, spin dynamics, and
transport properties in other channels.

\subsection{Description of the quasiparticle: Equation-of-motion method}

Now imagine a bare hole is injected into the ground state of $N_e$
electrons. By symmetry, such a state should be orthogonal to the ground
state of $N_e-1$ electrons. Its dissolution into a holon and a spinon is
also prohibited by the symmetry introduced by the phase shift field [(\ref
{mutual})] and the latter would otherwise cost a logarithmically divergent
energy if being left alone unscreened. Therefore, one has to treat a 
quasiparticle as an {\em independent} collective excitation in this spin-charge 
separation system.

Involving {\em infinite-body} holons and spinons, a quasiparticle cannot be
simply described by the mean-field theory of individual holon and spinon.
The previously discussed confinement is one example of {\em non-perturbative}
consequences caused by the infinite-body phase shift field. But such a
confinement of the holon and spinon inside a quasiparticle will enable us to
approach this problem from a different angle.

Here it may be instructive to recall how a low-lying collective mode is 
determined in the BCS  theory. In the BCS mean-field theory, quasiparticle 
excitations are well defined with an energy gap. But quasiparticle 
excitations do not exhaust all the low-lying excitations, and there exists a 
collective mode in the absence of long-range Coulomb interaction, which may be 
also regarded as a ``bound'' state of a quasiparticle pair due to the {\em 
residual} attractive interaction. A correct way\cite{eom} to handle this 
``bound'' state is to use the {\em full} BCS Hamiltonian to first write down the
equation of motion for a quasiparticle pair and {\em then} apply the BCS
mean-field treatment to linearize the equation to produce the gapless
spectrum, which is equivalent to the RPA scheme\cite{BCS}. Including the
long-range Coulomb interaction \cite{eom} will turn this collective mode
into the well-known plasma mode.

Similarly we can establish an equation-of-motion description of the 
quasiparticle as a ``collective mode'', which moves on the background of the 
mean-field spin-charge separation state. For
this purpose, let us first write down the {\it full} equation of motion of the
hole operator in the Heisenberg representation: $-i\partial_tc_{i%
\sigma}(t)=[H_{t-J}, c_{i\sigma}(t)]$, based on the {\em exact} $t-J$ model,
either in the decomposition (\ref{mutual}) or simply in the original $c$%
-operator representation, as follows 
\begin{eqnarray}  \label{ct}
[H_t, c_{i\sigma}]&=& \frac{t}{2}(1+n_i^h)\sum_{l=NN(i)}c_{l\sigma} 
\nonumber \\
&+& t\sum_{l=NN(i)}\left(c_{l\sigma} \sigma S^z_i+
c_{l-\sigma}S^{-\sigma}_i\right),
\end{eqnarray}
and 
\begin{eqnarray}  \label{cj}
[H_J, c_{i\sigma}]&=& \frac{J}{4} c_{i\sigma}\sum_{l=NN(i)} (1-n_l^h) 
\nonumber \\
&-& \frac{J}{2}\sum_{l=NN(i)}\left(c_{i\sigma} \sigma S^z_l+
c_{i-\sigma}S^{-\sigma}_l\right).
\end{eqnarray}
Note that the above equations hold in the restricted Hilbert space under the
no-double-occupancy constraint: $\sum_{\sigma}c^{\dagger}_{i\sigma}c_{i%
\sigma}\leq 1$.

There are many papers in literature dealing with the $t-J$
model in the $c$-operator representation, in which the no-double-occupancy $%
\sum_{i\sigma }c_{i\sigma }^{\dagger }c_{i\sigma }\leq 1$ is disregarded. As
aconsequence, there is only a conventional {\em scattering} between the
quasiparticle and spin fluctuations as suggested by (\ref{ct}) and (\ref{cj}%
). This leads to a typical spin-fluctuation theory, which usually remains a
Fermi liquid theory with well-defined coherent quasiparticle excitations
near the Fermi surface in contrast to the $Z=0$ conclusion obtained here.
The problem with the spin fluctuation theory is that the crucial role of the
no-double-occupancy constraint hidden in (\ref{ct}) and (\ref{cj}) has been
completely ignored which, in combination with the RVB spin pairing, is
actually the key reason resulting in a spin-charge separation state in the
present effective theory of the $t-J$ model. In such a backdrop of the
holon-spinon sea, the scattering terms in (\ref{ct}) and (\ref{cj}) will
actually produce a virtual ``decaying'' process which is fundamentally
different from the usual spin-fluctuation scattering in shaping the
single-electron propagator.

By using the decomposition (\ref{mutual}) and the mean-field order
parameter $\Delta^s$ defined in (\ref{deltas}), the high-order
spin-fluctuation-scattering terms on the r.h.s. of (\ref{ct})
and (\ref{cj}) can be ``reduced'' to the same order of linear $c_{i\sigma }$%
, and we find 
\begin{eqnarray}\label{cc}
-i\partial _tc_{i\sigma }(t) &\approx &\frac t2(1+\delta
)\sum_{l=NN(i)}c_{l\sigma }+J(1-\delta )c_{i\sigma }  \nonumber  
\\
&&-\frac 14tB_0\sum_{l=NN(i)}e^{i\hat{\Theta}_{l\sigma }}h_l^{\dagger
}b_{i\sigma }e^{-i\sigma A_{il}^h}+\frac 3 8J\Delta^s
\sum_{l=NN(i)}e^{i\hat{%
\Theta}_{i\sigma }}h_i^{\dagger }b_{l-\sigma }^{\dagger }e^{i\sigma
A_{il}^h}+...,
\end{eqnarray}
where $B_0$ is the modified (but not an independent) order parameter for the
hopping term introduced in Ref. \onlinecite{string3}. In the following we
will discuss some unique quasiparticle properties based on this equation.

So in the spin-charge separation (mean-field) background, the leading order
effect of the ``scattering'' terms correspond to the decay of the
quasiparticle: The terms in the second line of (\ref{cc}) clearly indicate
the tendency for the quasiparticle to break up into holon and spinon
constituents. This is in contrast to the conventional {\em scatterings}
between the quasiparticle and spin fluctuations, as (\ref{ct}) and (\ref{cj}%
) would have suggested. Generally, the quasiparticle is
expected to have an intrinsic broad spectral function extended over the whole
energy range 
\begin{equation}
E_{\mbox{quasiparticle}}>E_{\mbox{holon}}+E_{\mbox{spinon}}  \label{energy}
\end{equation}
because of the decomposition process. But the presence of the phase factor $%
e^{i\hat{\Theta}}$ in these ``decaying'' terms of (\ref{cc}) prevents a real
decay of the quasiparticle since such a vortex field would cost a
logarithmically divergent energy as has been discussed before. Thus, even in
the case of (\ref{energy}), the decaying of a quasiparticle remains only a
virtual process which is an another way to understand the confinement
discussed in Sec. II C. 

Without the ``decaying'' terms, the equation-of-motion (\ref{cc}) 
would become closed with an eigen spectrum in momentum-energy space 
(besides a constant which can be absorbed into the chemical potential): 
\begin{equation}
\epsilon _{{\bf k}}=-2t_{eff}(\cos k_x+\cos k_y),  \label{epsilon}
\end{equation}
with 
\begin{equation}
t_{eff}=\frac t2(1+\delta ).
\end{equation}
Generally the ``decaying'' terms do not contribute to a coherent 
${\bf k}$-dependent correction due to the nature of the holon and spinon 
excitations as well as the ``smearing'' caused by $e^{i\hat{\Theta}}$ in
(\ref{cc}).
But in the ground state, which is also superconducting, the ``decaying''
terms in (\ref{cc}) do produce a coherent contribution due to the composite
nature of the quasiparticle which will modify the solution of the
equation-of-motion.

\subsubsection{Ground state: a superconducting state}

In the mean-field ground state, the bosonic holons are Bose condensed with $%
\langle h_i^{\dagger}\rangle= h_0\sim \sqrt{\delta}$ and the superconducting order
parameter $\Delta^{SC}\neq 0$ (see Sec. II A). The decomposition (\ref
{mutual}) then implies that the electron $c$-operator may be rewritten in
two parts: 
\begin{equation}  \label{c}
c_{i\sigma}= h_0 {a}_{i\sigma}+ {c}_{i\sigma}^{\prime},
\end{equation}
where ${a}_{i\sigma}\equiv b_{i\sigma}e^{i\hat{\Theta}_{i\sigma}}$ and ${c}%
^{\prime}_{i\sigma}=(:h_i^{\dagger}:)b_{i\sigma} e^{i\hat{\Theta}_{i\sigma}}$
with $: h_i^{\dagger}:\equiv h_i^{\dagger}- h_0$. Correspondingly, a
coherent term will emerge from the ``decaying'' terms in (\ref{cc}) which is
linear in $a^{\dagger}$: 
\begin{eqnarray}
\mbox{$J$-scattering term in (\ref{cc})}\rightarrow \frac{3}{8}%
J\sum_{l=NN(i)}\left(\frac{\Delta_{il}^{SC}}{h_0^2}\right)\sigma h_0{a}%
_{l-\sigma}^{\dagger}+\mbox{high order}.
\end{eqnarray}
In obtaining the r.h.s. of the above expression, the superconducting order
parameter defined in (\ref{sc}) is used.

Note that the $t$-scattering term in (\ref{cc}) gives rise to a term $%
\propto h_0{a}_{i\sigma}$ which can be absorbed by the chemical potential $%
\mu$ added to the equation. Then one finds 
\begin{eqnarray}  \label{cbar}
-i\partial_t {a}_{i\sigma}&\simeq & t_{eff}\sum_{l=NN(i)}{a}_{l\sigma} + \mu 
{a}_{i\sigma} +\frac{3}{8}J\sum_{l=NN(i)}\left(\frac{\Delta_{il}^{SC}}{h_0^2}%
\right)\sigma {a}^{\dagger}_{l-\sigma} + \mbox{high order}
\end{eqnarray}
where the connection between $a$ and $c^{\prime}$ has been assumed to be in
high order and thus is neglected in the leading order approximation to get
a closed form in linear $a$ and $a^{\dagger}$. Finally, introducing the
Bogoliubov transformation in the momentum space 
\begin{equation}
a_{{\bf k} \sigma}=u_{{\bf k}} \gamma_{{\bf k}\sigma} -\sigma v_{{\bf k}}
\gamma^{\dagger}_{-{\bf k}-\sigma},
\end{equation}
we find that (\ref{cbar}) can be reduced to 
\begin{equation}
-i\partial_t \gamma_{{\bf k}\sigma}^{\dagger}=E_{{\bf k}} \gamma_{{\bf k}%
\sigma}^{\dagger},
\end{equation}
where $\gamma_{{\bf k}\sigma}^{\dagger}$ represents the creation operator of
an eigenstate of quasiparticle excitations with the energy spectrum 
\begin{equation}  \label{ek}
E_{{\bf k}}=\sqrt{(\epsilon_{{\bf k}}-\mu)^2+|\Delta_{{\bf k}}|^2}.
\end{equation}
Here $\Delta_{{\bf k}}$ is defined by 
\begin{equation}  \label{gap}
\Delta_{{\bf k}}=\frac 3 4 J\sum_{{\bf q}} \Gamma_{{\bf q}}\left( \frac{%
\Delta^{SC}_{{\bf k+q}}}{h_0^2}\right),
\end{equation}
with $\Gamma_{{\bf q}}=\cos q_x +\cos q_y $. Like in the BCS theory, $u_{%
{\bf k}}^2=[1+ (\epsilon_{{\bf k}}-\mu)/E_{{\bf k}}]/2$ and $v_{{\bf k}%
}^2=[1- (\epsilon_{{\bf k}}-\mu)/E_{{\bf k}}]/2$.

The large ``Fermi surface'' is defined by $\epsilon_{{\bf k}}=\mu$ and $%
\Delta_{{\bf k}}$ then represents the energy gap opened at the Fermi
surface. Note that $\Delta_{{\bf k}}$ changes sign as 
\begin{equation}
\Delta_{{\bf k+Q}}=-\Delta_{{\bf k}},
\end{equation}
with ${\bf Q}=(\pm \pi, \pm\pi)$, by noting $\Gamma_{{\bf q}+{\bf Q}%
}=-\Gamma_{{\bf q}}$ in (\ref{gap}). It means that $\Delta_{{\bf k}}$ has
opposite signs at ${\bf k}= (\pm \pi, 0)$ and $(0, \pm \pi_)$, indicating a
d-wave symmetry near the Fermi surface. In fact, since the pairing order
parameter $\Delta_{{\bf k}}^{SC}$ is d-wave-like\cite{string3}, $\Delta_{%
{\bf k}}$ should be always d-wave-like with node lines $k_x=\pm k_y$
according to (\ref{gap}).

Comparing to the conventional BCS theory with d-wave order parameter, there
are several distinct features in the present case. First of all, besides the
d-wave quasiparticle spectrum illustrated in Fig. 4 by the ``V'' shape lines
along the Fermi surface,  there exists a discrete spinon excitation level at 
$E_s\sim \delta J$ (horizontal line in Fig. 4) which leads to $E_g=2E_s\sim
41$$meV$ (if $J\sim 100$$meV$) magnetic peak at $\delta \sim 0.14$ as
reviewed in Sec. II A. This latter spin collective mode is {\it independent} of
the quasiparticle excitations at the mean-field level. 
\begin{figure}[ht!]
\epsfxsize=8.0 cm
\centerline{\epsffile{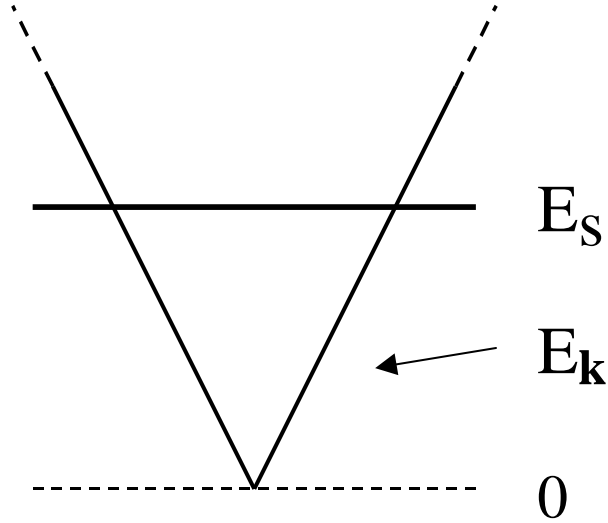}}
\vspace{2mm}
\caption{Low-lying excitations in the superconducting phase: The ``V"" 
shape quasiparticle spectrum and the discrete spinon energy at $E_s$. } 
\label{fig:4} 
\end{figure}

Secondly, even though the superconducting {\em order parameter} $%
\Delta^{SC}_{{\bf k}}$ and the {\em energy gap} $\Delta_{{\bf k}}$ in the
quasiparticle spectrum have the same symmetry: Both are d-wave-like, they
cannot be simply identified as the same quantity as in the BCS theory. For
example, while $\Delta^{SC}_{{\bf k}}$ apparently scales with the doping
concentration $\delta$ ($h_0 \propto \sqrt {\delta}$) and vanishes at $%
\delta \rightarrow 0$, the gap $\Delta_{{\bf k}}$ defined in (\ref{gap}) is
not, and can be {\em extrapolated} to a finite value in the zero doping limit
where $T_c=0$. It means 
\begin{equation}  \label{pred1}
\frac{2\Delta_{{\bf k}}(T=0)}{T_c}\rightarrow \infty
\end{equation}
at $\delta\rightarrow 0$, whereas the BCS theory predicts a constant $\sim
4.28$ (d-wave case\cite{zhu}). The result (\ref{pred1}) is consistent with
the ARPES measurements\cite{photo2}.

Thirdly, the quasiparticle gains a ``coherent'' part $h_0a$ which should 
behave similarly to the conventional quasiparticle in the BCS theory as it does
not further decay at $E_{\bf k}<E_s$ (see Fig. 4). In this sense, the 
quasiparticle partially restores its coherence in the
superconducting state. Such a coherent part will disappear as a result of
vanishing superfluid density. According to (\ref{c}) one may rewrite the
single-particle propagator as 
\begin{equation}
G_e\simeq h_0^2{G}_a+{G}_e^{\prime },  \label{g12}
\end{equation}
where $G_a$ denotes the propagator of $a$-particles with omitting the
crossing term between ${a}$ and ${c}^{\prime }$ which is assumed negligible.
Then $h_0^2{G}_a$ emerges as the ``coherent'' part of the Green's function
in superconducting state against the ``normal'' part ${G}_e^{\prime }$: 
\begin{equation}
h_0^2{G}_a({\bf k},\omega )\sim h_0^2\left( \frac{u_{{\bf k}}^2}{\omega -E_{%
{\bf k}}}+\frac{v_{{\bf k}}^2}{\omega +E_{{\bf k}}}\right) .  \label{barg}
\end{equation}
Correspondingly, the total spectral function as the imaginary part of $G_e$
in our theory can be written as 
\begin{equation}
A_e({\bf k},\omega )=h_0^2{A}_a({\bf k},\omega )+{A}_e^{\prime }({\bf k}%
,\omega ).
\end{equation}
So at $h_0\rightarrow 0$, even though $\Delta _{{\bf k}}$ does not scale
with $h_0$, the superconducting coherent part $h_0^2A_a$ vanishes
altogether, with $A_e$ reduces to the normal part $A_e^{\prime }$ at $T>T_c$.

Finally, in the present case, $A_e^{\prime}({\bf k}, \omega)$ as a normal
part has nothing to do with the procedure that leads to the spectrum (\ref
{ek}) with a d-wave gap, which is different from the slave-boson approach
where the fermionic spinons are all paired up such that even the part of the
spectral function corresponding to $A_e^{\prime}$ should also look like in a
d-wave pairing state. Due to the sum rule $\int \frac{d\omega}{2\pi} A^e(%
{\bf k}, \omega)=1$, the ``normal'' part ${A}_e^{\prime}({\bf k})$ is
expected to be sort of suppressed by the emergence of the ``coherent'' $h_0^2%
{A}_a$ part, but since the latter is in order of $\delta$, ${A}_e^{\prime}$
should be still dominant away from the Fermi surface at small doping. It
implies that even in superconducting state, a normal-state dispersion
represented by the peak of ${A}_e^{\prime}$ may still be present as a
``hump'' in the total spectral function $A_e$. Recent ARPES experiments have
indeed indicated\cite{photo3} the existence of a ``hump'' in the spectral
function which clearly exhibits normal-state dispersion in the Fermi-surface
portions near the areas of ($\pm\pi$, $0$) and ($0$, $\pm\pi$) where the
d-wave gap is maximum.

\subsubsection{Normal state}

In the normal state, without any coherent contribution, the scattering terms
in (\ref{cc}) only give rise to the virtual process for a quasiparticle to
decay into the holon-spinon pairs. Based on (\ref{cc}), the propagator can be
determined according to the following standard equation of motion for the
single-particle Green's function $G_e(i,j; t)$: 
\begin{equation}  \label{cg}
\partial_t G_e(i, j; t)=\theta(t)\langle
[H_{t-J},c_{i\sigma}(t)]c^{\dagger}_{j\sigma}(0)\rangle -\theta(-t)\langle
c^{\dagger}_{j\sigma}(0)[H_{t-J}, c_{i\sigma}(t)]\rangle-i\delta
(t)\delta_{i,j}.
\end{equation}
If we simply neglect the scattering terms in zero-order approximation, a
closed form for $G_e$ can be obtained in momentum-energy space: 
\begin{equation}  \label{gns}
G_e({\bf k}, \omega)\sim \frac {1}{\omega -( \epsilon_{{\bf k}}-\mu)}.
\end{equation}

Here the quasiparticle spectrum $\epsilon_{{\bf k}}$ [defined in (\ref
{epsilon})] is essentially the same as the original {\em band-structure}
spectrum except for a factor of $2/(1+\delta)\approx 2$ enhancement in
effective mass. It is noted that in the $t-J$ model if the hopping term
described by the tight-binding model is replaced by a more realistic
band-structure model, like introducing the next-nearest-neighbor hopping
terms, the above conclusion about the factor-2 enhancement of the effective
mass still holds, in good agreement with ARPES experiments\cite{photo}. Here
the reason for the mass enhancement is quite simple: At each step of
hopping, the probability is roughly one half for a hole not to change the
surrounding singlet spin configuration, which in turn reduces the
``bandwidth'' of the quasiparticle by a factor of two.

The expression (\ref{gns}) shows a ``quasiparticle'' peak at $\epsilon_{{\bf %
k}}- \mu$ and defines a large ``Fermi surface'' as an equal-energy contour
at $\epsilon_{{\bf k}}=\mu$. Here $\mu$ is determined such that $-i 2\sum_{%
{\bf k}} G_e({\bf k}, t=-0) = N_e$\cite{remark4}. So the ``Fermi surface''
structure should look like similar to that of a non-interacting
band-structure fermion system as long as the virtual decaying process in (%
\ref{cc}) does not fundamentally alter it. [As mentioned before, we do
not expect such ``decaying'' terms to significantly modify the 
${\bf k}$-dependence of $\epsilon_{\bf k}$ since, for example, the spinon
no longer has a well-defined spectrum in momentum space (see Sec. II A)
and in particular, the vortex phase $e^{i\hat{\Theta}}$ in (\ref{cc}) will
further ``smear out'' the ${\bf k}$-dependent correction, if any, from 
(\ref{cc}).]

So far we have not discussed the finite life-time effect of a
quasiparticle due to the ``decaying'' terms in (\ref{cc}). Even though the
true break-up of a quasiparticle is prevented by the phase shift field as
discussed before, the virtual decaying process should remain a very strong
effect since the phase shift field only cost a logarithmically divergent
energy at a large length scale. The corresponding confinement force is
rather weak and the virtual decaying process should become predominant locally
to cause an intrinsic broad feature in the spectral function at high energy.
Such a broad structure reflecting the decomposition in the one-hole case has 
been previously discussed in Ref. \onlinecite{1hole}. At finite doping, how the
``decaying'' effect shapes the broadening of the quasiparticle
peak will be a subject to be investigated elsewhere.

\subsubsection{Destruction of Fermi surface: Deconfinement of spinon and
holon}

The existence of a large Fermi surface, coinciding with the {\em %
non-interacting} band-structure one, can be attributed to the integrity of
the quasiparticle due to the confinement of spinon and holon. But as pointed
out in Sec. II C, such a confinement will disappear in the zero-doping
limit. The Fermi surface structure will then be drastically changed.

In this limit, the single-electron propagator may be expressed in the
following {\em decomposition} form 
\begin{equation}  \label{decomp}
G_e \approx i G_f\cdot G_b
\end{equation}
where 
\begin{equation}  \label{gh}
G_f(i, j; t) =-i\left\langle T_t h^{\dagger}_i(t)\left( e^{i\frac 1 2 {\Phi}%
^b_{i}(t)}e^{-i\frac 1 2 {\Phi}^b_{j}(0)}\right) h_j(0)\right\rangle,
\end{equation}
and 
\begin{equation}  \label{gb}
G_b (i,j;t)=-i(-\sigma)^{i-j}\left\langle T_t b_{i\sigma}(t) \left(e^{-i%
\frac{\sigma}{2}\Phi^h_i(t)}e^{i\frac {\sigma}{2}\Phi^h_j(0)}\right)b^{%
\dagger}_{j\sigma}(0)\right\rangle,
\end{equation}
without the multi-value problem because $\Phi^h_i$ in (\ref{phih}) vanishes
and $e^{i\frac 1 2 \Phi^b_i}$ becomes well-defined as discussed in Sec. II
C. At $\delta\rightarrow 0$, $H_s$ in (\ref{hs}) reduces to the SBMFT
Hamiltonian with $A^h=0$ and $G_b$ becomes the conventional Schwinger-boson
propagator. Such a deconfinement can be also seen from the equation of
motion (\ref{cc}) by noting that $e^{i\hat{\Theta}_{i\sigma}}\rightarrow e^{i%
\frac 1 2 \Phi^b_i}$ can be absorbed by $h^{\dagger}_i$, while $A^h_{il}=0$,
so that the scattering term becomes a pure decaying process for the
quasiparticle without any confining force. Due to such a true decaying,
equation (\ref{cc}) actually describes in real time the first step towards
dissolution for the quasiparticle. In particular, the large Fermi surface
structure originated from the {\em bare} hopping term in (\ref{cc}) will no
longer appear in the decomposition form of the electron propagator (\ref
{decomp}), where the residual Fermi surface (points) will solely come from
the oscillating part of the phase shift field $e^{i\frac 1 2 \Phi^b_i}$ in $%
G_f$.

The single-electron propagator for the one-hole case has been discussed in
detail in Ref.\onlinecite{1hole}. Here the large Fermi surface is gone
except for four {\em Fermi points} at ${\bf k}_0=(\pm\pi/2, \pm\pi/2)$ with
the rest part looking like all ``gapped''. In fact, in the convolution form
of (\ref{decomp}) the ``quasiparticle'' peak (edge) is essentially
determined by the spinon spectrum $E^s_{{\bf k}}=2.32J\sqrt{1-s_{%
{\bf k}}^2}$ with $s_{{\bf k}}=(\sin k_x+ \sin k_y)/2$ in SBMFT through the
spinon propagator $G_b$, since the holon propagator $G_f$ is incoherent\cite
{1hole}. The intrinsic broad feature of the spectral function found in Ref.%
\onlinecite{1hole} is due to the convolution law of (\ref{decomp}) and is a
direct indication of the composite nature of the quasiparticle, which is
also consistent with the ARPES results\cite{arpes1} as well as the earlier
theoretical discussion in Ref.\onlinecite{laughlin1}.

Note that the Fermi points ${\bf k}_0$ coming from $G_f$ at low energy is
due to the phase shift field $e^{i\frac 1 2 \Phi^b_i}$ appearing in it. In
Ref.\onlinecite{1hole}, this is shown in the slave-fermion formulation which
is related to the present formulation through a unitary transformation\cite
{string1} with $h_i^{\dagger}e^{i\frac 1 2 \Phi^b_i}$ being replaced by a new holon
operator $f_i^{\dagger}$. And the $f$-holon will then pick up a phase string factor $%
(-1)^{N_c^{\downarrow}}$ ($N_c^{\downarrow}$ denotes the total number of $%
\downarrow$ spins exchanged with the holon during its propagation along the
path $c$ connecting sites $i$ and $j$) at low energy which can be written as 
\begin{equation}
(-1)^{N_c^{\downarrow}}\equiv e^{\pm i \pi N_c^{\downarrow}}= e^{i{\bf k}%
_0\cdot ({\bf r}_i-{\bf r}_j)} e^{\pm i\delta N_c^{\downarrow}},
\end{equation}
where $\delta N_c^{\downarrow}= N_c^{\downarrow}-\langle N_c^{\downarrow}
\rangle$, and $\langle n^b_{l\downarrow}\rangle=1/2$ is used. On the other
hand, in the equal-time ($t\sim 0^-$) limit, the singular oscillating part
of $e^{iK_{\sigma}(c)}$ in (\ref{prop1}) will also contribute to a large
``remnant Fermi surface'' in the momentum distribution function $n({\bf k})$
which can be regarded as a precursor of the large Fermi surface in the
confining phase at finite doping, and is also consistent with the ARPES
experiment as discussed in Ref.\onlinecite{1hole}.

The above one-hole picture may have an important implication for the
so-called pseudo-gap phenomenon\cite{arpes2} in the underdoped region of the
high-$T_c$ cuprates. Even though the confinement will set in once the
density of holes becomes finite, the ``confining force'' should remain {\em %
weak} at small doping, and one expects the virtual ``decaying'' process in (%
\ref{cc}) to contribute significantly at weakly doping to bridge a
continuum evolution between the Fermi-point structure in the zero-doping
limit to a full large Fermi surface at larger doping. Recall that in the
one-hole case decaying into spinon-holon composite happens around ${\bf k}_0$
at zero energy transfer, while it costs {\em higher} energy near $(\pi, 0)$
and $(0,\pi)$, which shouldn't be changed much at weakly doping. In the
confinement regime, the quasiparticle peak in the electron spectral function
defines a quasiparticle spectrum and a large Fermi surface as discussed
before. Then due to the the virtual ``decaying'' process in the equation of
motion (\ref{cc}) [as shown in Fig. 1(b)], the spectral function will become
much broadened with its weight shifted toward higher energy like a
gap-opening near those portions of the Fermi surface far away from ${\bf k}_0
$, particularly around four corners $(\pm\pi, 0)$ and $(0,\pm\pi)$. With the
increase of doping concentration and reduction of the decaying effect, the
suppressed quasiparticle peak can be gradually recovered starting from the
inner parts of the Fermi surface towards four corners $(\pm\pi, 0)$ and $(0,
\pm\pi)$. Eventually, a coherent Landau quasiparticle may be even restored
in the so-called overdoped regime, when the bosonic RVB ordering collapses
such that the spin-charge separation disappears.

Furthermore, at small doping (underdoping), something more dramatic can
happen in the model described by (\ref{hh}) and (\ref{hs}). In Ref.%
\onlinecite{string3}, a microscopic type of {\em phase separation} has been
found in this regime which is characterized by the Bose condensation of
bosonic spinon field. Since spinons are presumably condensed in {\em %
hole-dilute} regions\cite{string3}, the propagator will then exhibit
features looking like in an even {\em weaker} doping concentration or more
``gap'' like than in a uniform case, below a characteristic temperature $T^*$
which determines this microscopic phase separation. Therefore, the ``spin
gap'' phenomenon related to the ARPES experiments\cite{arpes2} in the
underdoped cuprates may be understood as a ``partial'' deconfinement of
holon and spinon whose effect is ``amplified'' through a microscopic phase
separation in this weakly-doped regime. As discussed in Ref.%
\onlinecite{string3}, $T^*$ also characterizes other ``spin-gap'' properties
in magnetic and transport channels in this underdoping regime.

\section{CONCLUSION AND DISCUSSION: A Unified View}

In this paper, we have studied the quasiparticle properties of doped holes
based on an effective spin-charge separation theory of the $t-J$ model. The
most unique result is that a quasiparticle remains stable as an independent
excitation despite the existence of holon and spinon elementary excitations.
The underlying physics is that in order for a doped hole to evolve into
elementary excitations described by holon and spinon, the whole system has
to adjust itself globally which would take infinite time under a local
perturbation. Such an adjustment is characterized by a vortex-like phase
shift as shown in (\ref{mutual}). As a consequence of the phase shift
effect, the holon and spinon constituents are found to be effectively
confined which maintains the integrity of a quasiparticle except for the
case in the zero doping limit. In particular, the quasiparticle weight is
zero since there is no overlap between a doped hole state and the true
ground state due to the symmetry difference introduced by the vortex phase
shift. Such a quasiparticle is no longer a conventional Landau quasiparticle
and is generally incoherent due to the virtual decaying process. Only in the 
superconducting state the coherence can be partially regained by the 
quasiparticle excitation.

The physical origin of the ``unrenormalizable phase shift'' is based on the
fact that a hole moving on an antiferromagnetic spin background will always
pick up the phase string composed of a product of $+$ and $-$ signs which
depend on the spins exchanged with the hole during its propagation\cite
{string}. Such a phase string is nonrepairable at low energy and is the only
source to generate phase frustrations in the $t-J$ model. The phase shift
field in (\ref{mutual}) precisely keeps track of such a phase string 
effect\cite{string1} and therefore accurately describes the phase problem in 
the $t-J$ model even at the mean-field level discussed in Sec. II A.

Probably the best way to summarize the present work is to compare the
present self-consistent spin-charge separation theory with some fundamental
concepts and ideas proposed over years in literature related to the doped
Mott-Hubbard insulator.

{\em The RVB pairing.} The present theory can be regarded as {\em one} of
the RVB theories\cite{anderson,baskaran,anderson3}, where the spin RVB
pairing is the driving force behind everything from spin-charge separation to
superconductivity. The key justification for {\em this} RVB theory is that
it naturally recovers the {\em bosonic} RVB description at half-filling,
which represents\cite{liang,chen} the most accurate description of the
antiferromagnet for both short-range and long-range AF correlations. In the
metallic state at finite doping, the RVB order $\Delta^s$ defined in (\ref
{deltas}) reflects a partial ``fermionization'' from the original pure
bosonic RVB pairing due to the gauge field $A^h_{ij}$ determined by doped
holes. But it is still physically different from a {\em full} fermionic RVB
description\cite{anderson,zou,baskaran}. In contrast to the fermionic RVB
order parameter, $\Delta^s$ here serves as a ``super'' order parameter
characterizes a {\em unified} phase covering the antiferromagnetic
insulating and metallic phases, normal and superconducting states altogether%
\cite{string3}.

{\em Spin-charge separation.} In our theory, elementary excitations are
described by charge-neutral spinon and spinless holon fields, and the ground
state may be viewed as a spinon-holon sea. Different from slave-particle
decompositions, spinons and holons here are all {\em bosonic} in nature and
the conventional gauge symmetry is broken by the RVB ordering. But these
spinons and holons in 2D still couple to each other through the mutual
Chern-Simons-like gauge interactions which are crucial to $T_c$, anomalous
transport and magnetic properties. The Bose condensation of holons
corresponds to the superconducting state, while the Bose condensation of
spinons in the {\em insulating phase} gives rise to an AF long-range order.
The spinon Bose condensation can persist into the metallic regime, leading
to a pseudo-gap phase with microscopic phase separation which can coexist
with superconductivity\cite{string3}.

{\em Bosonization.} The electron c-operator expressed in terms of bosonic
spinon and holon in (\ref{mutual}) naturally realizes a special form of
bosonization. A 2D bosonization description has been regarded by many\cite
{anderson3,haldane,bosonization1,bosonization2,bosonization3} as the
long-sought technique to replace the perturbative many-body theory in
dealing with a non-Fermi liquid. The 2D bosonization scheme has been usually
studied, as an analog to the successful 1D version\cite{haldane1}, in the
momentum space where Fermi surface patches have to be assumed first\cite
{anderson3,haldane,bosonization1,bosonization2,bosonization3}. In the
present scheme, which is also applicable to 1D, the Fermi surface satisfying
the Luttinger volume and the so-called Fermi-surface fluctuations are
presumably all {\em generated} by the phase shift field in (\ref{mutual}),
which guarantees the fermionic nature of the electron. Note that the
vortex-structure involved in the phase shift field in the 2D case is the
main distinction from the $\Theta$-function in conventional bosonization
proposals \cite{bosonization3}.

{\em Non-Fermi liquid.} As a consequence of the phase shift field,
representing the Fermi surface ``fluctuations'', the ground state is a
non-Fermi liquid with the vanishing spectral weight $Z$, consistent with the
argument made by Anderson\cite{anderson2,anderson3} based on a
``scattering'' phase-shift description. In 1D both methods are equivalent as
the phase shift value in the latter can be determined quantitatively based
on the exact Bethe-Ansatz solution. But in 2D, the phase shift field $%
\Theta^{string}_{i\sigma}$, which is obtained by keeping track of the
nonrepairable phase string effect induced by the traveling holon, provides a
unique many-body version with vorticities, and our model shows how a {\em %
concrete} 2D non-Fermi liquid system can be realized.

{\em Quasiparticle: Spinon-holon confinement.} In conventional spin-charge
separation theories based on slave-particle schemes, a quasiparticle does
not exist at all: It always breaks up and decays into spinon-holon
elementary excitations. This deconfinement has been widely perceived as a
logical consequence of the spin-charge separation in literature. But in the
present paper, we have shown that the phase shift field actually {\em %
confines} the spinon-holon constituents (at least at finite doping), which
means the integrity of a quasiparticle is still preserved even in a
spin-charge separation state. It may be considered as a $U(1)$ version of
quark confinement, but with a twist: the stable quasiparticle as a
collective mode is generally not a {\em coherent} elementary excitation
since during its propagation the phase shift field also induces a nonlocal
phase string on its path. In other words, the quasiparticle here is not a
Landau quasiparticle anymore. The confinement and nonrepairable phase string
effect both reflect the fermionic nature of the quasiparticle, namely, the 
{\em fermionic} quasiparticle cannot simply decay into {\em bosonic} spinon
and holon; and the phase string effect comes from sequential signs due to
the {\em exchange} between the propagating quasiparticle and the
spinon-holon background --- the latter after all is composed of fermionic
electrons in the original representation.

The issue of the possible confinement of spinon and holon was already raised
by Laughlin\cite{laughlin2} along a different line of reasoning. He has also
discussed numerical and experimental evidence that the spinon-holon may be
seen, more sensibly, in high-energy spectroscopy like the way quarks are
seen in particle physics. In the recent SU(2) gauge theory\cite{su(2)}, an
attraction between spinon and holon to form a bound state due to gauge
fluctuations is also assumed in order to explain the ARPES data. But in the
present work the essential point is that the holons and spinons are not
confined in the ground state but are only bound in quasiparticles as a kind
of incoherent (many-body) excitations which have no overlap with elementary
holon and spinon excitations as guaranteed by symmetry. These incoherent
quasiparticle excitations should not have any significant contribution to
the thermodynamic properties (at least above $T_c$). At short distance and
high-energy, the composite nature of a quasiparticle will become dominant
which may well explain the broad {\em intrinsic} structure in the spectral
function observed in the ARPES experiments. The composite structure of the
quasiparticle can even show up at low-energy when one navigates through
different circumstances like the superconducting condensation, underdoping
regime, etc., with some unique features different from the behavior expected
from Landau quasiparticles.

{\em Fractional statistics.} Laughlin\cite{laughlin3} had made a compelling
argument right after the proposal of the spin-charge separation that the
holon and spinon should carry fractional statistics, by making an analogy of
the spin liquid state with a fractional quantum Hall state. Even though the
absence of the time-reversal symmetry-breaking evidence in experiment does
not support the original version of fractional statistics (anyon) theories%
\cite{laughlin3,wilczek}, the essential characterization of fractional
statistics for the singlet spin liquid state is, surprisingly, present in
our theory in the form of the phase string effect as discussed in Ref.%
\onlinecite{string1}. But no explicit time-reversal symmetry is broken in
this description\cite{string3}.

A fractional statistics may sound strange as we have been talking about
``bosonization'' throughout the paper. But as pointed out in Ref.%
\onlinecite{string1}, if the phase shift field $\Theta_{i\sigma}^{string}$
is to be ``absorbed'' by the bosonic spinon and holon fields, then the
expression (\ref{mutual}) can be regarded as a slave-semion decomposition
with the new ``spinon'' and ``holon'' fields being ``bosonic'' among
themselves but satisfying a {\em mutual} fractional statistics between the
spin and charge degrees of freedom. The origin of mutual statistics can be
traced back to the nonrepairable phase string effect induced by a hole in
the antiferromagnetic spin background. At finite doping, the order parameter 
$\Delta^s$ in (\ref{deltas}) actually describes the RVB pairing of spinons
with {\em mutual-statistics} which reduces to the bosonic RVB only in the
half-filling limit. Furthermore, in the bosonic representation of (\ref{hh})
and (\ref{hs}) the lattice Chern-Simons fields, $A_{ij}^f$ and $A_{ij}^h$,
precisely keep track of mutual statistics\cite{string1}, which are crucial
to various peculiar properties exhibited in the model.

{\em Gauge theory.} Based on the slave-boson decomposition, it has been shown%
\cite{gauge1,gauge2} that the gauge coupling is the most important
low-energy interaction associated with spin-charge deconfinement there. The
anomalous linear-temperature resistivity\cite{gauge2} has become the
hallmark for anomalous transport phenomenon based on the scattering between
charge carriers and gauge fluctuations. In the present theory, the holons
are also subject to strong random flux fluctuations, in terms of the
effective holon Hamiltonian (\ref{hs}), in a uniform normal state where it
leads\cite{weng2} to the linear-T resistivity in consistency with the Monte
Carlo numerical calculation\cite{MC}. Other anomalous transport properties
related to the cuprates may be also systematically explained in such a
simple gauge model based on some effective analytic treatment\cite
{weng2,weng3}. In contrast to the slave-boson gauge theory, however, the
spinon part does not participate in the transport phenomenon due to the RVB
condensation ($\Delta^s\neq 0$) persisting over the normal state, and
besides the Chern-Simons gauge fields $A_{ij}^{f, h}$, the conventional
gauge interaction between spinons and holons is suppressed because of it.

Furthermore, the $\pi$-flux phase\cite{marston} and commensurate flux 
phase\cite{comm} in the mean-field
slave-boson theory, and the recent SU(2) gauge theory\cite{su(2)} at small doping
have a very close connection with the present approach: A {\em fermionic}
spinon in the presence of $\pi$ flux per plaquette is actually a precursor
to become a {\em bosonic} one at half-filling under the lattice and
no-double-occupancy constraint. In Ref.\onlinecite{weng2}, how such a
statistics-transmutation occurs has been discussed, and in fact the
bosonization decomposition (\ref{mutual}) was first obtained there based on
the fermionic flux phase. In spite of physical proximity, however, the
detailed mathematical structure of gauge description\cite{gauge3,gauge4} for
the fermionic flux phase and the present bosonic spinon description are
obviously rather different.

{\em Superconducting mechanism.} Anderson\cite{anderson} originally
conjectured that the superconducting condensation may occur once the RVB
spin pairs in the insulating phase start to move like Cooper pairs in the
doped case. The superconducting condensation in the present theory indeed
follows suit. But there is an important subtlety here. Since the RVB paring
order parameter $\Delta^s$ covers the normal state as well, there must be an
another factor controlling the superconducting transition: The {\em phase
coherence}. Indeed, the vortex phase $\Phi^b$ appearing in the
superconducting order parameter (\ref{sc}) is the key to ensure the {\em %
phase coherence} at a relatively low temperature compatible to a
characteristic {\em spin} energy\cite{string3,string4}. In other words, the
phase coherence discussed by Emery and Kivelson\cite{emery} is realized by
the {\em phase shift field} in the present theory which effectively resolves
the issue why $T_c$ is too high in previous RVB theories.

Furthermore, the interlayer pair tunneling mechanism\cite{ilayer} for
superconductivity is also relevant to the present theory from a different
angle. Recall that the quasiparticle does exist in the present theory but is
always {\em incoherent} just like the blocking of a coherent single-particle
interlayer hopping conjectured in Ref.\onlinecite{ilayer}. On the other
hand, a pair of quasiparticles in the singlet channel can recover the {\em %
coherency} due to the cancellation of the frustration caused by the phase
string effect. Thus if one is to construct a phenomenological theory based
on the {\em electron} representation, the superconductivity can be naturally
viewed as due to an {\em in-plane} kinetic mechanism\cite{baskaran1}.

{\em Phase string and $Z_2$ gauge theory} As pointed out at the beginning of
this section, our whole theory is built on the phase string effect
identified in the $t-J$ model. Namely, the main phase frustration induced by
doping is characterized by a sequence of signs $\prod_c \sigma_{ij}$ on a
closed path $c$ where $\sigma_{ij}=\pm 1$ denotes the index of spins
exchanged with a hole at a link $(ij)$ during its hopping. Thus, instead of
working in the vortex representation of (\ref{mutual}) where the singular
phase string effect has been built into the wave functions with the
non-singular part described by the Chern-Simons-like lattice gauge fields%
\cite{string1}, one may also directly construct a 2+1 $Z_2$ gauge theroy\cite{z2} to deal with the singular phase string $\prod_c \sigma_{ij}$. A
discrete $Z_2$ gauge theory here seems the most natural description of the
phase string effect as the sole phase frustration in the $t-J$ model, in 
contrast to the conventional continuum gauge field 
description\cite{gauge1,gauge2}.

Finally, we emphasize the close connection between the antiferromagnetism
and superconductivity as both occur in a unified RVB background controlled
by $\Delta ^s$. The relation between the AF insulating phase and the
superconducting phase here is much more intrinsic than in conventional
approaches to the $t-J$ model. Especially, the coexistence of holon and
spinon Bose condensations in the underdoped regime\cite{string3} makes a
group theory description of such a phase, in a fashion of the SO(5) theory%
\cite{so5}, become possible but with an important modification:
Inhomogeneity must play a crucial role here in the Bose condensed holon and
spinon fields in order to describe this underdoped regime. Detailed
investigation along this line will be pursued elsewhere.

\acknowledgments

The present work is supported, in part, by the Texas ARP program No. 3652707
and a grant from the Robert A. Welch foundation, and the State of Texas
through the Texas Center for Superconductivity at University of Houston.


\begin{references}
\bibitem{remark0}  Strictly speaking, this orthogonal condition holds only
exactly in the thermodynamic limit. A finite-size lattice usually breaks the
continuum rotational symmetry. In the thermodynamic limit, when the length
scale under consideration is much larger than the lattice constant, i.e., in
the continuum limit, a rotational symmetry and angular momentum can still be
meaningfully defined. Since $l$ is a thermodynamic number [$\sim O(N)$] at
finite $\delta $, it will still remain so while the local discrete
coordinates get coarse-grained in the continuum limit.
\bibitem{remark4}  Note that $\mu $ here is an independent variable which
has not previously appeared in the mean-field theory of the ground state
discussed in Sec. II A where only the Lagrangian multiplier $\lambda $ is
present to enforce the double-occupancy constraint at global level while the
holon number is fixed\cite{string3}.
\end{references}
\end{document}